\documentclass[%
reprint,
superscriptaddress,
 amsmath,amssymb,
 aps,
floatfix
]{revtex4-2}

\usepackage{graphicx}
\usepackage{xcolor}
\usepackage{dcolumn}
\usepackage{bm}

\graphicspath{{figures/}}

\newcommand{\SLAC}{SLAC National Accelerator Laboratory, Menlo Park, California 94025, USA}
\newcommand{\StanfordApplPhys}{Department of Applied Physics, Stanford University, Stanford, California 94305, USA}
\newcommand{\Sandia}{Sandia National Laboratories, Albuquerque, New Mexico 87123, USA}
\newcommand{\StanfordMech}{Department of Mechanical Engineering, Stanford University, Stanford, California 94305, USA}
\newcommand{\TelAviv}{The School of Physics and Astronomy, Tel-Aviv University, Tel-Aviv, 69978, Israel}

\newcommand{\LMITAU}{Tel-Aviv University Center for Light-Matter-Interaction, Tel-Aviv 69978, Israel}
\newcommand{\TAUeng}{The Faculty of Engineering, Tel-Aviv University, Tel-Aviv 69978, Israel}
\begin{document}

\title{Continuous Laser-Driven $\alpha$ Source}

\author{N. Popper}
\affiliation{\TelAviv}
\affiliation{\LMITAU}
\author{G. D. Glenn}
\affiliation{\SLAC}
\affiliation{\StanfordApplPhys}
\affiliation{\Sandia}
\author{G. Jain}
\affiliation{\SLAC}
\affiliation{\StanfordMech}
\author{R. Halifa Levi}
\affiliation{\SLAC}
\affiliation{\TAUeng}
\author{T. Catabi}
\affiliation{\TelAviv}
\affiliation{\LMITAU}
\author{M. Elkind}
\affiliation{\TelAviv}
\affiliation{\LMITAU}
\author{A. Levinson}
\affiliation{\TelAviv}
\affiliation{\LMITAU}
\author{A. Levanon}
\affiliation{\TelAviv}
\affiliation{\LMITAU}
\author{S. H. Glenzer}
\affiliation{\SLAC}
\author{I. Pomerantz}
\email{ipom@tauex.tau.ac.il}
\affiliation{\TelAviv}
\affiliation{\LMITAU}

\date{\today}

\begin{abstract}
Laser-driven proton-boron fusion offers a path for generation of $\alpha$ particles on a compact scale.  Conventional solid-target experiments, however, are fundamentally limited by target degradation and low shot-repetition capacity. Here, we demonstrate a continuous laser-driven $\alpha$ particle source utilizing target-normal sheath acceleration  from a continuously replenishable ultrathin water-leaf jet target. The proton beam interacts with a downstream boron nitride converter, driving the $^{11}\text{B}(p,\alpha)^{8}\text{Be}$ reaction which yields three $\alpha$ particles.
The measurement was repeated across  $18{,}000$ consecutive shots at a rate of $10\,\text{Hz}$. The outgoing $\alpha$-particle yield and spectral fluence were measured and benchmarked against comprehensive particle-transport simulations. The numerical and experimental results reveal a pronounced flattening in the energy-differential $\alpha$-particle spectrum escaping from the converter, driven by depth-dependent reaction yield and stopping power effects. The sustained stable $\alpha$-particle production at high cumulative yields paves the way for applications in targeted particle therapy and laser-driven plasma-fusion research.
\end{abstract}

\maketitle

\section{Introduction}
\label{sec:intro}

Fusion energy holds the transformative potential to provide a virtually limitless, clean, and safe baseload power supply for the future. 
Among the various candidate fuel cycles, p--$^{11}$B fusion is  attractive due to its predominantly aneutronic nature; the reaction proceeds through $^{11}\mathrm{B}(p,\alpha){}^{8}\mathrm{Be}$, followed by the prompt breakup of the resonance state $^{8}\mathrm{Be}$ into two additional $\alpha$ particles.
The successive nuclear interactions yield low neutron doses relative to other fusion fuels, thereby relaxing the requirement for thick fusion-blanket systems.
However, these  advantages are counterbalanced by significant physical disadvantages, primarily the temperatures required for ignition that are much higher than those of D-T fuel, and the enormous bremsstrahlung radiation losses that make achieving a self-sustaining thermonuclear burn difficult.

The p--$^{11}$B reaction cross section peaks at CM energy of $\sim$0.6~MeV, readily achieved with pico- and femto-second intense-laser acceleration schemes.
Laser-driven p--$^{11}$B fusion has been studied experimentally for over two decades \cite{belyaev2005observation}. Modern experimental schemes typically employ either an ``in-target'' geometry \cite{giuffrida2020,istokskaia2023,molloy2025}, where homogeneous composite targets or layered nanomaterials are directly irradiated, or a ``pitcher-catcher'' configuration, in which a high-intensity laser accelerates a proton beam that subsequently impacts a secondary boron target \cite{margarone2020generation,sciscio2025laser,schollmeier2022investigation}. 

Experimental venues explored to increase fusion yields include harnessing the produced energy of secondary nuclear reactions in the target \cite{hora2015fusion}, and relaxing the limits imposed by thermal equilibrium 
by sustaining the reaction inside laser-generated boron plasma \cite{labaune2013fusion}.

Beyond energy production, laser-driven p--$^{11}$B reaction offers  potential for targeted $\alpha$-particle therapy \cite{jelinek2024first}.
However, despite rapid advancements and continuously increasing reaction yields, the absolute $\alpha$-particle doses produced to date remain too low for the realization of these applications.

In this work we demonstrate continuous $\alpha$ generation using laser-driven target-normal-sheath-acceleration (TNSA) of protons \cite{passoni2010target},
realized through a replenishable ultrathin water-leaf jet target  \cite{Crissman2022,he2025stable}.

\section{Numerical experiments}
\label{sec:numerical}
Using the FLUKA particle-transport code \cite{bohlen2014fluka}, we first studied numerically the production of $\alpha$ particles when an MeV-level proton beam impinges on a BN target. The simulation accounts for the energy loss of the incident protons in the target, the energy dependence of the $^{11}\mathrm{B}(p,\alpha){}^{8}\mathrm{Be}$ cross section, the energy distribution of the produced $\alpha$ particles, and finally the transport of $\alpha$ particles through the target. The source proton beam characteristics follow those of the experiment that will be described in the following section. The energy spectrum and the angular spread of the proton beam were parametrized according to the measured data shown in Fig.~\ref{fig:proton-stability}. A $1.2~\mathrm{cm} (x)\times~1.7~\mathrm{cm} (y)\times~1.7~\mathrm{cm} (z)$ BN cuboid  was positioned $7~\mathrm{cm}$ downstream of the proton source.

\begin{figure}
\centering
\includegraphics[width=65mm]{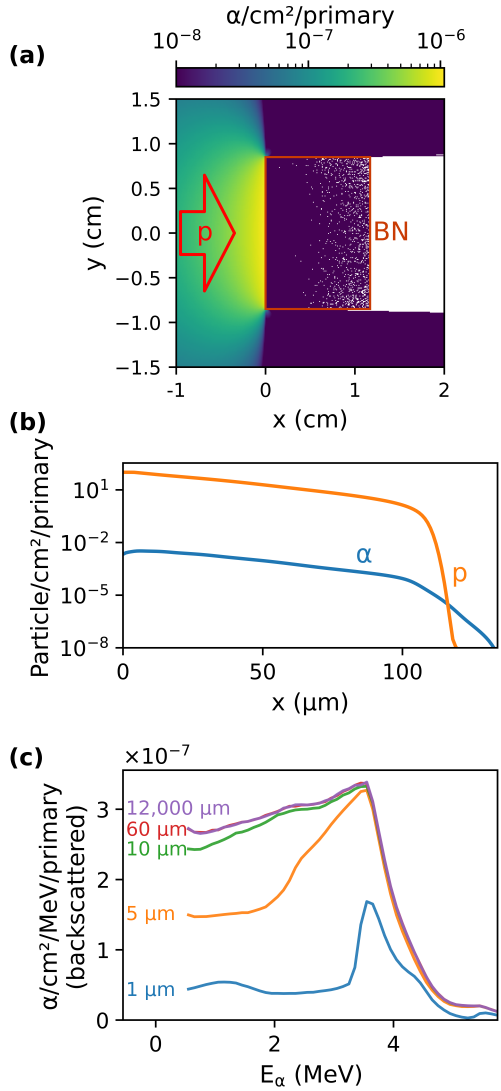}
\caption{
Particle transport simulation of a proton beam impinging on a BN target. (a) The fluence of the produced $\alpha$  particles. (b) The fluence of $\alpha$ particles and protons averaged over the transverse plane, plotted as a function of the depth in the target. (c) The fluence of $\alpha$ particles backscattering from the BN, for different target thicknesses. The thickest target (12,000 $\mu$m) corresponds to the one used in the experiment. 
}
\label{fig:numerical}
\end{figure}

Fig.~\ref{fig:numerical}(a) shows a 2D fluence distribution of the produced $\alpha$ particles. The proton beam impinges on the BN target along the $+x$ direction. Note that the incident beam cross section is larger than the BN target's face. The figure shows that the majority of $\alpha$ generation occurs in a thin layer at the front, where most of the outgoing $\alpha$ particles are observed to be backscattered. $\alpha$ particle generation deeper in the target is driven by neutrons from $^{11}$B(p,n)$^{11}$C reactions that subsequently induce $^{10}$B(n,$\alpha$)$^{7}$Li reactions. 

A zoomed-in view of this thin layer is shown in Fig.~\ref{fig:numerical}(b), where the  fluences of the $\alpha$ particles and protons are averaged over the transverse plane. The proton attenuation is consistent with the range of 2.5~MeV protons in BN, found to be 57~$\mu$m using the SRIM code \cite{Ziegler2010}. 

The simulated $\alpha$-to-proton fluence ratio is consistent with the expected $^{11}\mathrm{B}(p,\alpha){}^{8}\mathrm{Be}\rightarrow3\alpha$ reaction rate. At $E_p\simeq2.5~\mathrm{MeV}$, the evaluated cross section is $\sigma_\simeq350~\mathrm{mb}$ \cite{zerkin2018experimental}, 
while natural BN with a density of $2.1~\mathrm{g~cm^{-3}}$ has $n_{^{11}\mathrm B}\simeq4.08\times10^{22}~\mathrm{cm^{-3}}$. 
The range of $2.9~\mathrm{MeV}$ $\alpha$ particles (p--$^{11}$B reaction average) in BN is approximately $\overline{L}=9.1~\mu\mathrm{m}$, based on the measured energy-loss rate of $0.32~\mathrm{MeV\mu m^{-1}}$~ \cite{doan2017response}. The expected track-length fluence ratio is therefore $3n_{^{11}\mathrm B}\sigma\overline{L}\simeq3.9\times10^{-5}$,
in close agreement with the observed ratio in Fig.~\ref{fig:numerical}(b).

Fig.~\ref{fig:numerical}(c) shows the energy-differential fluence of $\alpha$s escaping from the target in the backscattered direction. Each curve corresponds to a different simulation run, where only the target thickness was varied in the range of 1~$\mu$m to 12,000~$\mu$m. The latter value corresponds to the target thickness in the experiment. The  $\alpha$ dose is observed to be saturated at target thicknesses larger than $\sim10\,\mu$m, consistent with the range of the escaping  $\alpha$ particles.

Fig.~\ref{fig:numerical}(c) also shows a systematic change in spectral shape with target thickness. For the thinnest target (1~$\mu$m), the escaping spectrum retains the structure of the emission spectrum, with a peak near 4~MeV. As the thickness increases this structure is progressively filled in, and once the target exceeds the proton range the spectrum becomes nearly featureless.
The mechanism is source self-absorption, well established in $\alpha$ spectroscopy of thick radioactive sources \cite{kadel2016analytic}. $\alpha$-particles reaching the surface at a given energy originate from a distribution of depths, and lower detected energies draw on emission from a thicker slab. The growth in contributing volume toward lower energies compensates the flux lost to stopping, leaving dN/dE approximately constant over much of the escaping spectrum. Because the $\alpha$ escape range in BN is only $\sim9\,\mu$m, this depth-averaging is confined to a thin surface layer even though $\alpha$ production extends over the full $\sim60\,\mu$m proton range.

\section{Experimental setup and irradiation geometry}
\label{sec:setup}

\begin{figure}
\centering
\includegraphics[width=0.9 \linewidth]{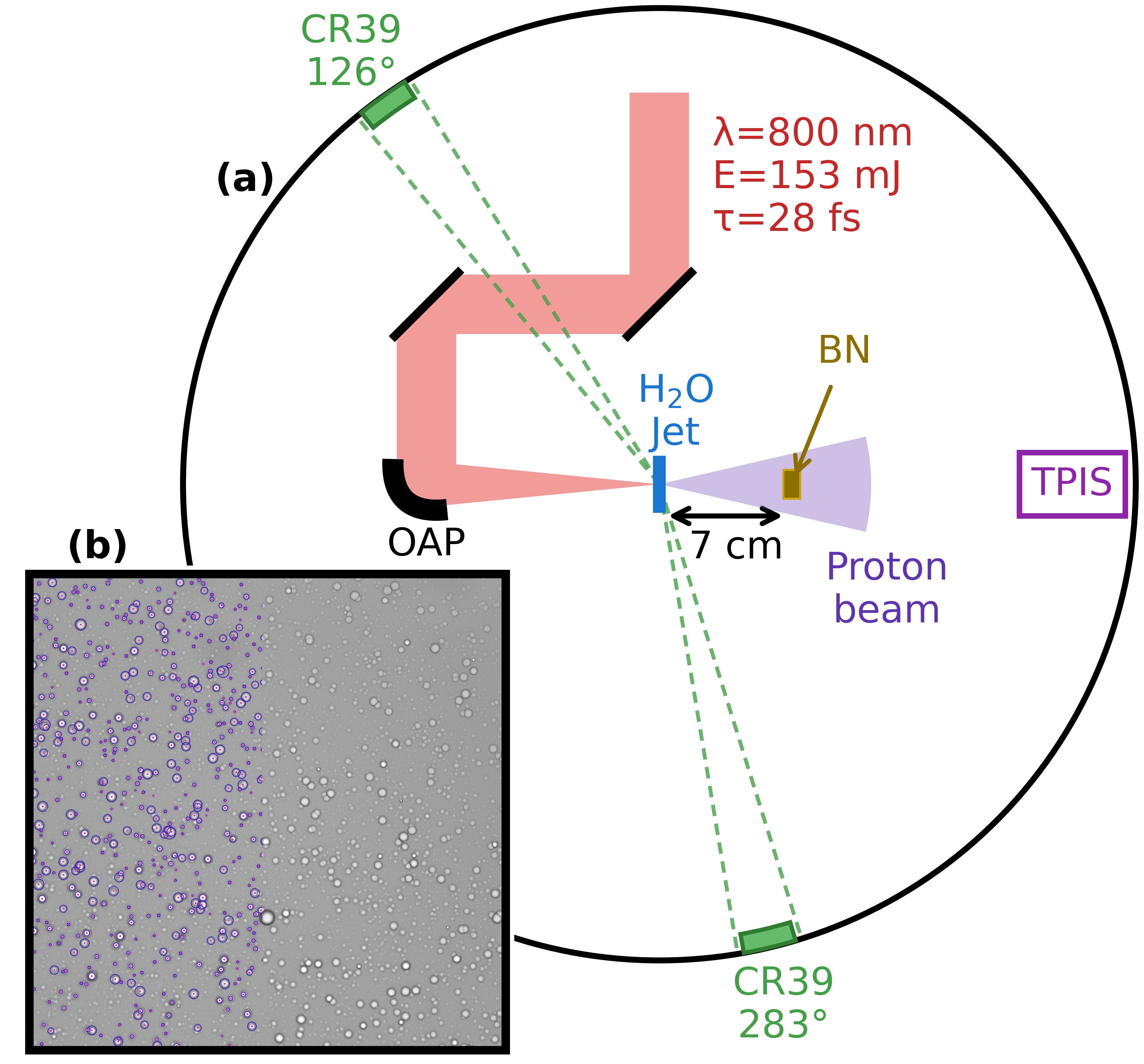}
\caption{(a) Schematics of irradiation setup.
The laser is focused by a f/2.5 off-axis parabolic mirror (OAP) to the water-leaf target. Ions emitted from the target impinge on the boron nitride (BN) converter. The resulting $\alpha$ particles are measured by CR39 nuclear track detectors. In different shots, a Thomson parabola ion spectrometer (TPIS) was used to characterize the ion beam. (b) Example image of a CR39 scan. 
The right part of the image shows the raw scan only, while
the left part is overlaid with circles obtained using the pit detection algorithm.}
\label{fig:system-schematic}
\end{figure}

We used the 20-TW NePTUN laser system at Tel Aviv University \cite{Porat2019} to generate proton beams at a rate of 10 Hz from a continuous flowing water-leaf target. 
These proton beams generated $\alpha$ particles via the $^{11}\mathrm{B}(p,\alpha){}^{8}\mathrm{Be}$ reaction in a BN converter. The experimental setup is shown schematically in Fig.~\ref{fig:system-schematic}.  28-fs long laser pulses of central wavelength $\lambda$~=~800~nm, with energy of ~153~mJ (on-target) and pulse contrast better than $10^{11}$ before $t = -60$~ps \cite{PhysRevResearch.3.L032059} were focused using an f/2.5 off-axis parabolic mirror onto the liquid sheet target, which was 0.38~µm-thick at the interaction point.
The radius containing 70\% of the energy was measured to be 3.1~µm, corresponding to a Gaussian-equivalent waist of w$_0$~=~4.0 µm, and a normalized laser amplitude of $a_0$~=~3.19.

A continuously replenishing liquid-H\textsubscript{2}O-sheet target was produced by a glass microfluidic converging nozzle (Micronit model Micro 1)~\cite{Crissman2022,Hoffman2022a,Treffert2022a,Faubel2025}. 
The nozzle formed a stable, approximately 2×0.5 mm$^2$
 sheet with a position-dependent micrometer-to-sub-micrometer thickness, allowing the target thickness to be selected by adjusting the laser-interaction position. The liquid was intercepted by a heated catcher located 4~mm below the interaction point and continuously removed and recirculated using an automated valve system. This configuration provided a debris-free target that was fully replenished between successive laser shots. 

\begin{figure}
\centering
\includegraphics[width=0.9 \linewidth]{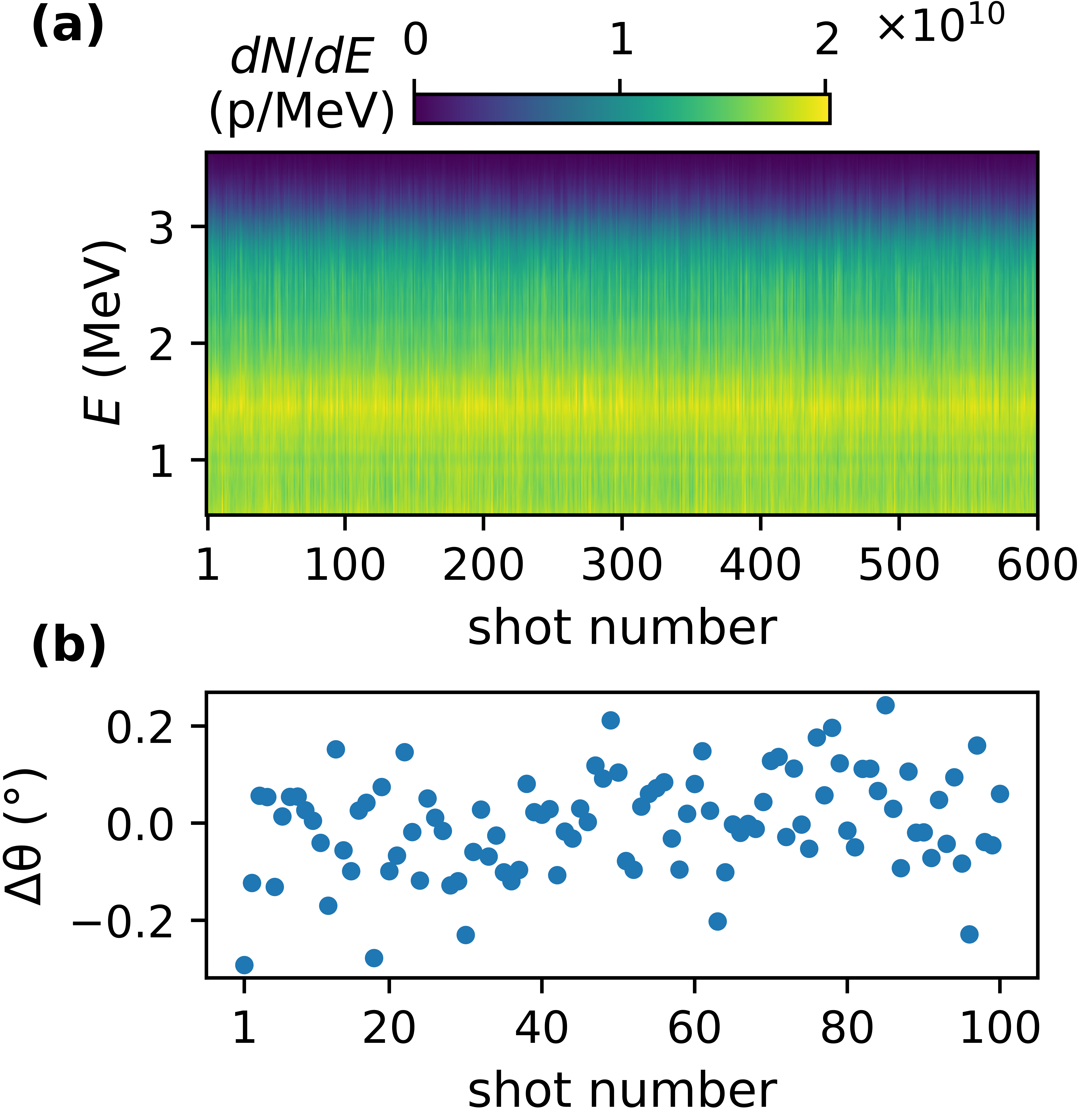}
\caption{ Stability of the proton source. (a) Energy spectra from 600 shots.  These data were averaged and used as the input spectrum for the FLUKA simulation. (b) Pointing stability of the proton beam from 100 shots.}
\label{fig:proton-stability}
\end{figure}

The proton beam was characterized by a Thomson parabola ion spectrometer (TPIS) which is described in detail in Ref. \cite{elkind2025ion}.
Example measured proton beam spectra, taken over 600 shots during a period of 60 seconds are shown in Fig.~\ref{fig:proton-stability}(a).
These results feature a proton cut-off energy of 3.33~MeV with a 0.23~MeV stability (RMS).

To measure the proton cone angle, a slotted scintillator (Mitsubishi Chemical Co. Ltd., DRZ-High) with an 11 $\mu$m Al filter was used and imaged. The slot was aligned to the TPIS path. The beam was characterized as an angular Gaussian with $\sigma = 0.22\,\mathrm{rad}$, comparable to TNSA proton beam divergences reported in the literature \cite{poole2018}.

The proton beam spatial profile was measured by filtering the image for a contiguous bright feature using standard morphological operations and determining the best-fit ellipse to this feature. The center of the ellipse determined the ion beam pointing direction. The overall beam divergence was calculated from the geometric mean of the semimajor and semiminor axes, using the scintillator distance from the interaction point to convert this to an angle. 
The proton-beam pointing spread over 100 consecutive shots is shown in  Fig.~\ref{fig:proton-stability}(b). The spread is found to be 0.10$^\circ$ (RMS). This corresponds to displacement much smaller than the transverse dimensions of the BN converter, confirming that the effect of pointing variations on $\alpha$ particle production is negligible.

A cuboid-shaped BN converter with dimensions of 1.2~cm (laser direction) and 1.7~cm $\times$ 1.7~cm (transverse directions), was placed 7~cm downstream of the water jet target. The proton cone overfills the  front face of the converter.

The resulting $\alpha$ dose was measured by two sets of CR39 nuclear-track detectors \cite{rana2018cr}. These detectors were placed at the interaction plane 
at 283$^\circ$ and 126$^\circ$ with respect to the laser direction, at a distance of 44~cm from the BN converter, each covering a 7.85$^\circ$ arc.

A prerequisite for obtaining absolutely calibrated spectral information from the pit size distribution in CR39, is absolute control over the etching conditions: concentration and temperature of the solution, and the etching time. 
For this reason, our CR39 detectors were treated at the External Dosimetry Laboratory located at Soreq Nuclear Research Center and measured using an automated microscopy system (Landauer Neutrak \cite{amit2020performance}).
The pit sizes were calibrated using $\alpha$s with known energies in the range of 0.5--5.4~MeV \cite{amit_cr39_alpha_spectroscopy_2024}.

We obtained the pit size distribution of each etched CR39 detector by applying a Hough circle transform  to identify candidate tracks, and then fitted a circular form to extract each pit size. This procedure was tuned on a single dataset and then applied blindly to the remaining sets.

\section{Results}
\label{sec:results}
\begin{figure*}[!htb]
\centering
\includegraphics[width=0.8\textwidth]{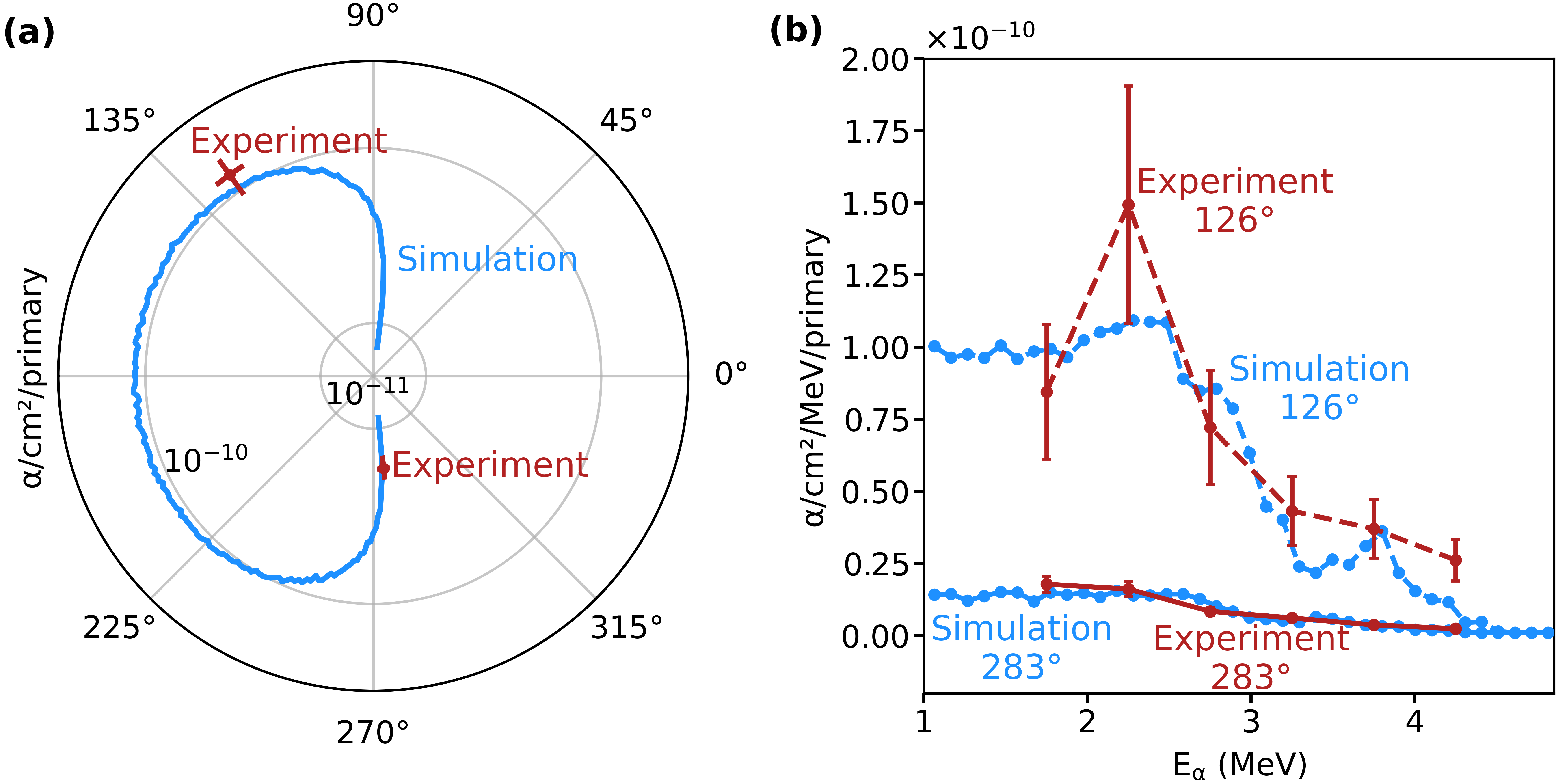}
\caption{$\alpha$ fluence per primary proton: experimental measurement using CR39 (red) vs. simulation (blue). (a) The total fluence. The indicated detector angles are with respect to the water-leaf target, i.e. the proton source. (b) Energy-differential fluences. The uncertainty bars represent the systematic error resulting from the pit-size identification procedure.}
\label{fig:results}
\end{figure*}

To benefit from the large proton dose enabled by the laser's 10 Hz repetition-rate and the continuous water-leaf target, we conducted a 1,800 s irradiation experiment. The resulting measured fluences of $\alpha$ particles are plotted in Fig.~\ref{fig:results}(a) in red. 
To remove low-energy particle background, a minimum pit-size cut of 6~µm was imposed,  corresponding to an $\alpha$ energy of 1.4~MeV.
The values are normalized to the total number of generated protons. The  error bars represent the angular size of the detector and the systematic uncertainty resulting from the pit-counting  procedure described above. The latter was retrieved by computing the standard deviation of the count between images of neighboring small areas on the same CR39 detector. 

The simulated $\alpha$ fluences per primary proton are overlaid on the figure in blue.
Orthogonal-distance regression gives discrepancies of $2.1\sigma$ and $1.9\sigma$, for the 283$^{\circ}$ and 126$^{\circ}$ detectors, respectively, where $\sigma$ is the combined experimental and simulation uncertainty.

Fig.~\ref{fig:results}(b) presents energy-differential fluences for the same experiments. The  flattening of the spectrum discussed above, the spectral fall-off at high energies, and the overall magnitude of the $\alpha$ doses are all recreated by the simulation. 
The larger fluence uncertainty in the 126$^\circ$ direction results from the higher density of pits which sometimes overlap, making them more difficult to resolve during image processing. 

The total number of pits measured at the 126$^\circ$ detector
corresponds to 4.9$\times$10$^8$ $\alpha$-particles/sr. 
Extrapolating according to the angular distribution obtained from simulation 
(Fig.~\ref{fig:results}(a)), this corresponds to 3.1$\times$10$^9$ $\alpha$-particles in total,
or 1.7$\times$10$^5$ per laser-shot.

\section{Conclusions}
Targets that handle both the primary laser interaction and the secondary $\alpha$ conversion (in-target) have been optimized \cite{molloy2025} and applied 
to high-repetition-rate $\alpha$ generation \cite{istokskaia2023}. 
These target discs, however, are of finite size, and therefore exhaustible.  

We demonstrated a particle generator that decouples proton generation from the boron converter and thus enables sustained, repeatable $\alpha$ production, at an average rate of 1.7$\times$10$^6$ $\alpha$/s.
A continuously replenished water-leaf target enabled utilizing the full 10~Hz repetition rate of the laser across  1800~s of irradiation, forming an accumulated $\alpha$ dose that is limited by run-time only.

The high yields of our experimental approach enabled detailed measurement of
the flattening of the escaping $\alpha$ spectrum.
Resolving this effect in laser-based p--$^{11}$B experiments is important because the measured distribution is the result of broadband proton production, depth-dependent reaction yield, and energy-dependent $\alpha$ stopping in the boron-containing converter. 
This effect is  central for future laser-driven p--B plasma schemes \cite{batani2023,Cirrone_Consoli_et_al_2025}, where using boron in plasma state to increase the escape fraction is a primary goal.

The water leaf target introduces two major hurdles that must be addressed for reliable operation. First, the increased water vapor load may require either physical separation of the vacuum systems with a window or a robust differential pumping arrangement; otherwise, the vapor could damage the compressor optics and degrade both the boron converter and particle transport performance. 
Second, the water leaf also raises the risk of laser back-reflection into the laser system, since gradual changes in leaf flatness can alter the optical return path; this makes careful tuning of the water-jet parameters essential to maintain stability and protect the laser.

We presented a laser-based experimental platform that enabled stable and sustained $\alpha$ particle production at a high repetition rate. Future work can extend this platform to higher laser intensities and repetition rates, as well as to more advanced boron-containing converters, including plasma-state targets that may increase the $\alpha$ escape fraction.  

\begin{acknowledgments}
We gratefully acknowledge D. P. DePonte and C. Hampton from the SLAC Sample Environment \& Delivery team and the Tel-Aviv University Mechanical Workshop for Research and Development for timely assistance during our experiment. G.D.G., G.J.,  and S.H.G. acknowledge support from the U.S. DOE Office of Science, Fusion Energy Sciences under FWP Nos.~100182 and 100866, and in part from the NSF under Grant Nos.~1632708 and PHY-2308860. G.D.G. acknowledges support from the DOE NNSA SSGF program under DE-NA0003960. 
S.H.G. acknowledges support from the Fulbright Scholar program, sponsored by the U.S. Department of State's Bureau of Educational and Cultural Affairs and USIEF.
Sandia National Laboratories is a multimission laboratory managed and operated by the National Technology and Engineering Solutions of Sandia, LLC, a wholly owned subsidiary of Honeywell International, Inc., for the U.S. Department of Energy’s National Nuclear Security Administration under Contract No. DE-NA0003525.
This paper describes objective technical results and analysis. Any subjective views or opinions that might be expressed in the paper do not necessarily represent the views of the U.S. Department of Energy or the United States Government.
This study was supported by grant no. 2022767
from the United States-Israel Binational Science Foundation (BSF)
and by the Israeli Ministry of Energy, grant no. 221-11-049.
\end{acknowledgments} 

\section*{Data Availability Statement}
The data that support the findings of this study are available from the corresponding author upon reasonable request.

\bibliography{mainNotes}

\end{document}